\documentclass[12pt]{article}
\usepackage{myown,epsfig}
\newcommand{\gam}{$\gamma$}
\newcommand{\etal}{{\it et al}}
\title{\large \sf \begin{flushright} Report 
     NPI \v{R}e\v{z}--TH--02/2000
    \end{flushright}
\vspace{5mm}
 \bf Internal conversion 
     coefficients for superheavy elements}
\author{\normalsize O. Dragoun, M.Ry\v sav\'y\thanks{e-mail: 
     rysavy@ujf.cas.cz}
   \ and  A.\v Spalek \\
\small \it Nuclear Physics Institute, Acad. Sci. of Czech Republic,
    \vspace{-0.7ex}\\
\small \it CZ-250 68 \v Re\v z near Prague, Czech Republic}
\date{ }
\begin{document}

\maketitle
\title{}
\begin{abstract}
The internal conversion coefficients (ICC) were calculated for all
atomic subshells of the elements with 104$\leq$Z$\leq$126, 
the E1...E4, M1...M4 multipolarities and the transition energies 
between 10 and 1000 keV. The atomic screening was treated in the
 relativistic Hartree-Fock-Slater model. The tables comprising 
almost 90000 subshell and total ICC were deposited at LANL preprint 
server \cite{Ry00}.
\end{abstract}
{\bf PACS}: 23.20.Nx, 27.90.+b\\
\section{Introduction}
   The \gam--ray and conversion electron spectroscopies have
contributed substantially to our knowledge of the nuclear decay
schemes. The extensive tables of the theoretical internal conversion
coefficients (ICC) have enabled the investigators to derive from
the spectroscopic data the multipolarities and total intensities of the
electromagnetic transitions depopulating the excited nuclear states.
Existing tables of the ICC cover wide region of the atomic numbers:
30$\leq$Z$\leq$104 for all atomic shells \cite{Roe78} and 10$\leq$Z$\leq$104
for the K, L, and M shells \cite{Ba78}.

   Very recently, Reiter \etal\ \cite{Rei99} investigated the Z=102 isotope
$^{254}$No produced in a heavy ion reaction. By means of a multidetector
\gam--ray spectrometer, the authors identified the ground-state band
of the even--even nucleus up to spin 14. The transitions of energy less
than 100 keV could not be seen in the \gam--ray spectrum since they
proceed almost entirely via emission of the conversion electrons. In
connection with a great effort to produce superheavy (SH) nuclei, e.g.
the indication for formation of the Z=118, A=293 isotope \cite{Nin99},
a further progress in the \gam--ray and conversion electron
spectroscopies can be expected. With this motivation, we calculated
in this work the ICC  for SH elements with Z up to 126.
\section{Calculations}
The internal conversion coefficients for the elements with Z $\geq$ 104
were evaluated using the program NICC \cite{Ry77}. The physical
model used for the atom description was the Hartree--Fock one with
the Slater's exchange term, the nucleus was described as the finite one
with the Fermi distribution of charge. With the exception of 
the nucleus description, this model is identical with that utilized by
R\"{o}sel \etal\ \cite{Roe78} (they used a homogenously charged
sphere). The electron wavefunctions -- both for the bound and free
electrons -- were calculated as solutions of the Dirac equation
with the atomic potential of Lu \etal\ \cite{Lu71} using the 
formulae of B\"{u}hring \cite{Bue65}. The conversion matrix elements
were then evaluated by direct integration with reasonably small
step. The resulting ICC correspond to the first non-vanishing
order of the perturbation theory.

The kinetic energy of the converted electron is very well approximated by
the difference
between the transition energy and the binding energy of this electron
before the conversion. Since the electron binding energies are
not known for the SH atoms, we used the eigenvalues
from \cite{Lu71}. There are two reasons for it. First, the eigenvalues
\cite{Lu71} approximate the binding energies fairly well (up to tens
of eV for the K--shell, up to eV's at the outermost shells) for
the lower Z. Second, the theoretical ICC are not too sensitive 
to small changes of the electron kinetic energy (except for the transition
energies very close, say hundreds of eV, to the threshold).
\section{Results and discussion}
We have calculated the ICC for all subshells of the elements
with 104 $\leq$ Z $\leq$ 126, the multipolarities E1 -- E4 and M1 -- M4,
and 16 energies from 10 keV up to 1 MeV. Altogether, almost 90000
coefficients were evaluated. The total ICC's were determined
using the subshell occupation numbers of \cite{Lu71}.

\hspace*{-\parindent}
\parbox{11cm}{\epsfxsize=9cm \epsfysize=7cm \epsfbox[20 100 330 400]
{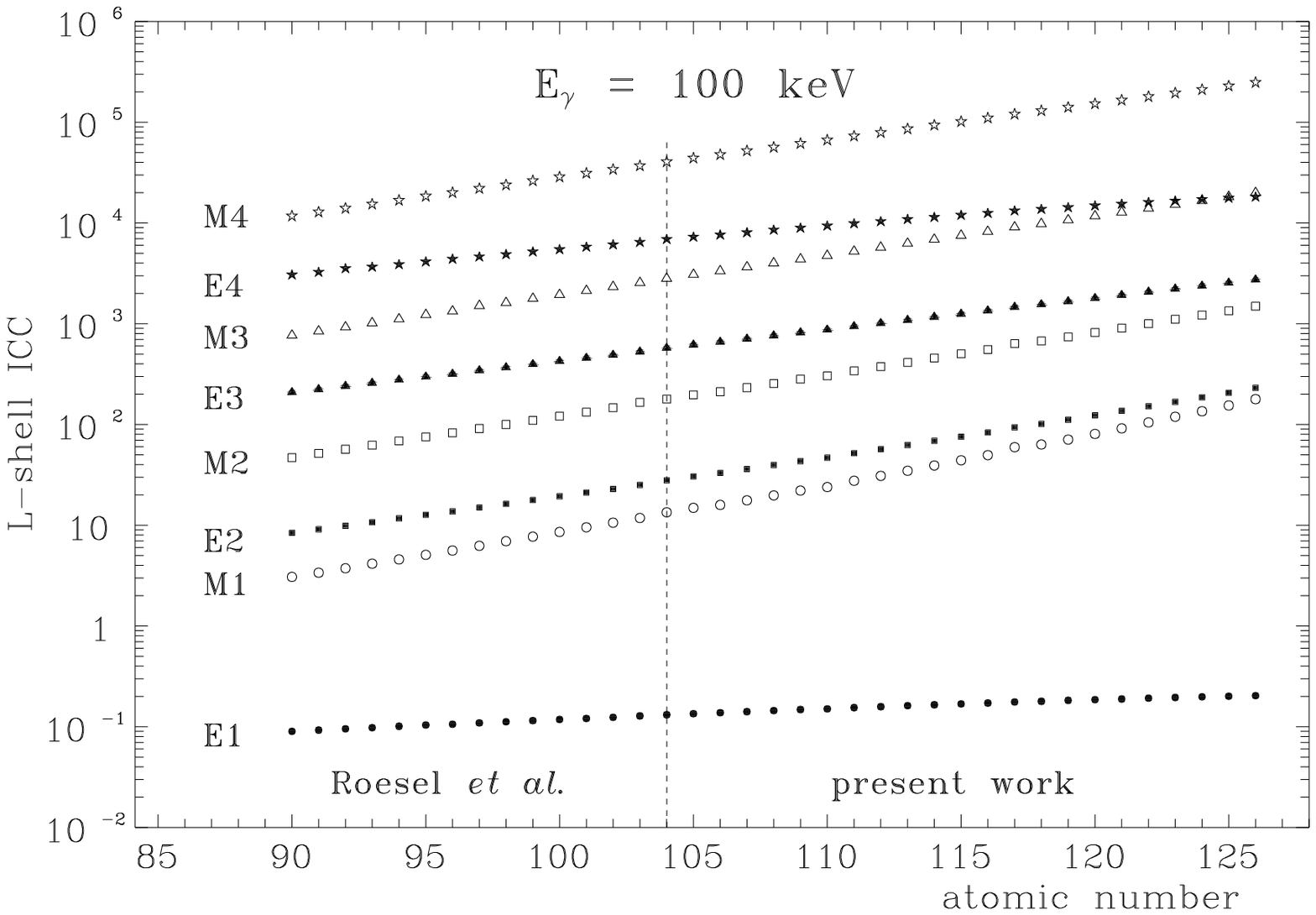}} \vspace{4ex}
\begin{figure}[h]
\protect\caption[fig1]{Examples of the dependence of ICC on atomic number, Z, for
   higher Z. Up to Z=103, the coefficients of R\"{o}sel et al.
   \cite{Roe78} are used while our results are plotted for Z$\geq$104.
   (Note that for Z=104, the ICC of \cite{Roe78} agree
   with our ones within 1--2 \%.)}
\label{f:Z_dep}
\end{figure}

We did not notice any unusual behaviour of the subshell ICC in
the very high Z region. Generally, the subshell ICC are increasing with
Z and the multipolarity, and decreasing with the transition energy.
Non-monotonous energy dependence can appear near the threshold.
As exemplified in figure \ref{f:Z_dep}, the ICC of this work represent a smooth
extention of the tables of Roesel \etal\ \cite{Roe78} for Z$>$104.
 For Z=104,
we compared the two sets of the ICC for various subshells and
various transition energies and multipolarities. In 100 comparisons,
the ICC of reference \cite{Roe78}
and those of this work agreed typically
within 1 - 2 \%. In order to show the important role of the internal
conversion in the region of superheavy elements, we present in
table \ref{t:tab1} the total ICC for every
 fifth Z value and a wide scope of
transition energies and multipolarities.
\begin{table}[h]
\begin{center}
\caption{Total internal conversion coefficients for selected
superheavy elements.}
 \label{t:tab1}
\begin{tabular}{rllllllll}
\hline
\multicolumn{1}{l}{E$_\gamma$} & \multicolumn{8}{c}{Multipolarity}\\
\cline{2-9}
$[$keV$]$   & E1 & E2 & E3 & E4 & M1 & M2 & M3 & M4 \\
\hline 
\multicolumn{9}{c}{Z=105} \\
\hline 
  30 &  2.36(0) & 8.04(3) & 6.26(5) & 3.93(7)&  1.81(2)&  3.65(4)&
  8.36(6) & 5.59(8)\\
 100  & 1.82(-1) & 4.31(1) & 9.53(2) & 1.28(4) & 2.01(1)&  2.79(2)&
  4.84(3) & 7.53(4)\\
 200 &  1.25(-1)&  2.09(0)&  2.48(1) & 1.98(2)&  1.13(1)&  3.68(1)&
  1.34(2) & 8.30(2)\\
 500 &  2.05(-2)&  1.08(-1) & 4.80(-1) & 1.81(0) & 8.82(-1)&  1.92(0)&
  3.89(0) & 9.03(0)\\
1000  & 6.43(-3)&  2.49(-2) & 6.74(-2) & 1.55(-1)&  1.36(-1)&  2.74(-1)&
  4.47(-1) & 7.44(-1) \\
\hline 
\multicolumn{9}{c}{Z=110} \\
\hline 
  30 &  2.58(0)&  1.21(4)&  8.95(5) & 5.87(7) & 2.97(2) & 5.87(4)&
  1.39(7)&  9.30(8)\\
 100  & 2.05(-1)&  6.69(1)&  1.38(3) & 1.75(4)&  3.24(1) & 4.36(2)&
  7.61(3)&  1.18(5)\\
 200 &  1.33(-1) & 3.21(0)&  3.70(1) & 2.83(2) & 1.70(1) & 5.13(1) &
 1.88(2)&  1.24(3)\\
 500  & 2.35(-2)&  1.52(-1)&  7.01(-1) & 2.65(0) & 1.30(0)&  2.72(0)&
  5.27(0)&  1.21(1)\\
1000 &  7.62(-3)&  3.33(-2)&  9.29(-2) & 2.16(-1) & 1.96(-1) &
  3.90(-1)&  6.14(-1) & 9.97(-1)\\
\hline 
\multicolumn{9}{c}{Z=115} \\
\hline 
  30 &  2.81(0) & 1.89(4) & 1.31(6) & 8.88(7) & 5.59(2) & 9.99(4) &
2.33(7) & 1.56(9)\\
 100 &  2.30(-1) & 1.09(2) & 2.05(3) & 2.42(4) & 6.01(1) & 7.32(2) &
1.23(4) & 1.87(5)\\
 200 &  4.58(-2) & 5.22(0) & 5.76(1) & 4.17(2) & 8.07(0) & 4.23(1) &
2.71(2) & 1.91(3)\\
 500 &  2.71(-2) & 2.33(-1) & 1.09(0) & 4.07(0) & 2.19(0) & 4.14(0) &
7.58(0) & 1.72(1)\\
1000 &  9.41(-3) & 4.88(-2) & 1.39(-1) & 3.22(-1) & 3.22(-1) & 6.07(-1) &
9.04(-1) & 1.43(0)\\
\hline 
\multicolumn{9}{c}{Z=120} \\
\hline 
  30 &  1.33(0) & 1.68(4) & 1.46(6) & 1.32(8) & 1.04(3) & 1.02(5) &
1.63(7) & 1.64(9)\\
 100 &  2.55(-1) & 1.80(2) & 3.03(3) & 3.34(4) & 1.10(2) & 1.21(3) &
1.97(4) & 2.94(5)\\
 200 &  5.32(-2) & 8.81(0) & 8.97(1) & 6.10(2) & 1.47(1) & 7.00(1) &
4.28(2) & 2.93(3)\\
 500 &  3.10(-2) & 3.67(-1) & 1.72(0) & 6.28(0) & 3.64(0) & 6.24(0) &
1.08(1) & 2.43(1)\\
1000 &  1.14(-2) & 7.24(-2) & 2.09(-1) & 4.82(-1) & 5.22(-1) & 9.30(-1) &
1.31(0) & 2.01(0)\\
\hline 
\multicolumn{9}{c}{Z=125} \\
\hline 
  30 &  1.40(0) & 2.87(4) & 2.32(6) & 2.03(8) & 2.03(3) & 1.78(5) &
 2.79(7) & 2.82(9)\\
 100 &  2.77(-1) & 3.06(2) & 4.53(3) & 4.59(4) & 2.12(2) & 2.02(3) &
3.16(4) & 4.60(5)\\
 200 &  6.13(-2) & 1.53(1) & 1.42(2) & 8.95(2) & 2.80(1) & 1.18(2) &
6.77(2) & 4.49(3)\\
 500 &  3.53(-2) & 6.06(-1) & 2.79(0) & 9.86(0) & 6.26(0) & 9.57(0) &
1.56(1) & 3.48(1)\\
1000 &  1.39(-2) & 1.12(-1) & 3.24(-1) & 7.42(-1) & 8.73(-1) & 1.46(0) &
1.93(0) & 2.87(0)\\
\hline 
\end{tabular}
\end{center}
\end{table}

     An example of the energy dependence of the total ICC is
depicted in figure \ref{f:zuby}. For each atomic subshell, there is
a discontinuity at its treshold energy. Note that not all
of them are seen in the logarithmic scale of figure \ref{f:zuby}, 
in dependence
on the relative contribution of the particular subshell to the total
ICC. For other multipolarities, the subshells with another angular
momenta can play more important role and the most  pronounced
discontinuity will appear at another energy (e.g. at the L$_1$ threshold
for the M1 multipolarity and the L$_2$ and L$_3$ thresholds for the E2 one).

\hspace*{-\parindent}
\parbox{11cm}{\epsfxsize=10cm \epsfysize=6cm \epsfbox[-30 100 330 400]
{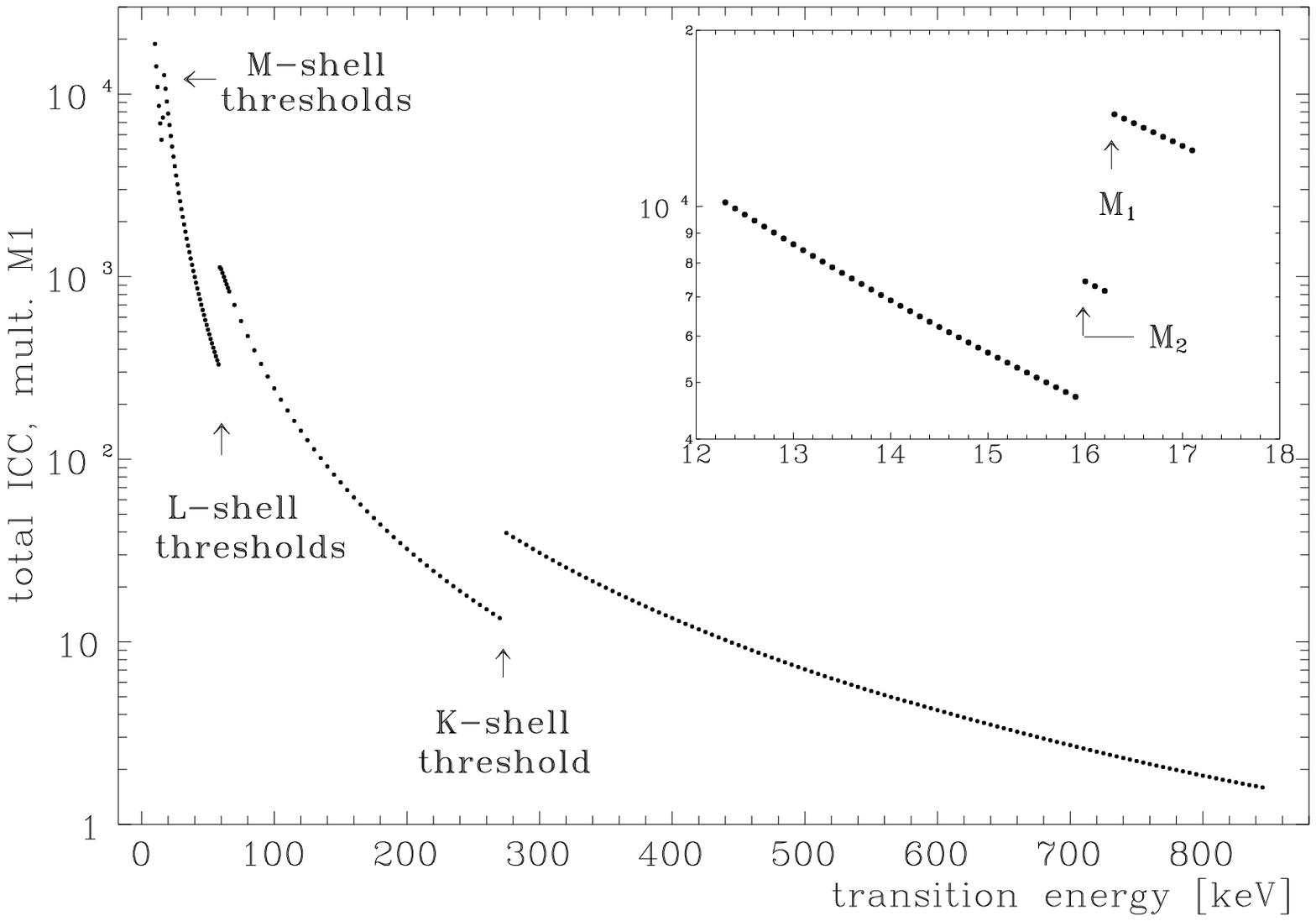}} \vspace{4ex}
\begin{figure}[h]
\protect\caption[fig2]{Example of dependence of the total ICC 
on the transition
   energy. The case of Z=126, multipolarity M1. The function is
   piecewise monotonous (decreasing) with discontinuities at the points
   where the transition energy reaches the binding energy of some
   subshell --- see text. In the inlet, the situation in the 
   neighbourhood of the M$_1$ and M$_2$ binding energies
   is depicted in more detail.}
\label{f:zuby}
\end{figure} 

    Figure \ref{f:zuby} also demonstrates why it is not 
reasonable to interpolate
in the tables of the total ICC if the transition energy is not
far enough above the K-shell threshold.  If it is not possible
to evaluate the ICC for the particular energy directly, the only
more or less reliable way is to
interpolate in the tables of the subshell ICC. We have to remind that
the systematic errors unfortunately appeared in the tables of the
{\it total}
ICC of Roesel \etal\ \cite{Roe78} as explained in our previous work
\cite{Dra92}.
(We also noticed that otherwise useful graphs of the ICC, presented
in Appendix F1 of widely distributed Table of Isotopes \cite{Fi96} show
incorrect behaviour of the {\it total} ICC in the neighbourhood of some
energy thresholds.)

Considering the important role of the internal conversion we have also
inspected the line shape which could be observed in the electron 
spectra in experiments with the SH nuclei.
 The measured 
spectra are influenced by energy losses and scattering of 
conversion electrons emerging 
from the target where they originated.

	To estimate this effect we performed the Monte Carlo 
calculations (see, e.g., \cite{Dra99} for description) of 
electron scattering for the conditions typical in the experiments 
with SH nuclei. We 
considered the case of 600 nm thick lead target and the initial 
electron energy of 10 keV. The 
isotropic initial distribution of emitted conversion electrons 
was assumed apart from a 
possible angular dependence of electron emission in the case 
of nuclei produced in the 
reaction. Assuming prompt electron emission from the created 
SH atoms, the 
homogeneous distribution of electron sources in the target 
volume was used. The total number of the simulated initial 
electrons was 10 millions.
The energy distribution of all electrons emitted
into the 2$\pi$ solid angle corresponding to the space behind
the plane of the target surface was calculated. The curves 
corresponding to the convolution of the Gaussian spectrometer 
response function with the MC calculated distribution are
displayed in figure \ref{f:tonda}. 
The Gaussian curves with the FWHM = 0.5 and 1~keV  were used.
  It is seen that in this case of a
realistic instrumental resolution of the present semiconductor
spectrometers the electron scattering in the 
target causes increase of the 
spectral line width but it does not hinder the observation of interesting 
internal conversion 
lines. Moreover, with increasing electron energy and for thinner targets 
the effect of electron 
energy losses and scattering will be smaller. In addition, the SH
atoms move after the 
recoil in nuclear reaction and therefore some conversion electrons can 
be emitted in the sites 
closer to the target surface or even outside of the target.

	The more comprehensive Monte Carlo study of the electron 
scattering effect in the 
experiments with the SH nuclei will be presented 
at the conference MC2000 in Lisbon.
\hspace*{-\parindent}
\parbox{11cm}{\epsfxsize=10cm \epsfysize=6cm \epsfbox[-30 100 330 400]
{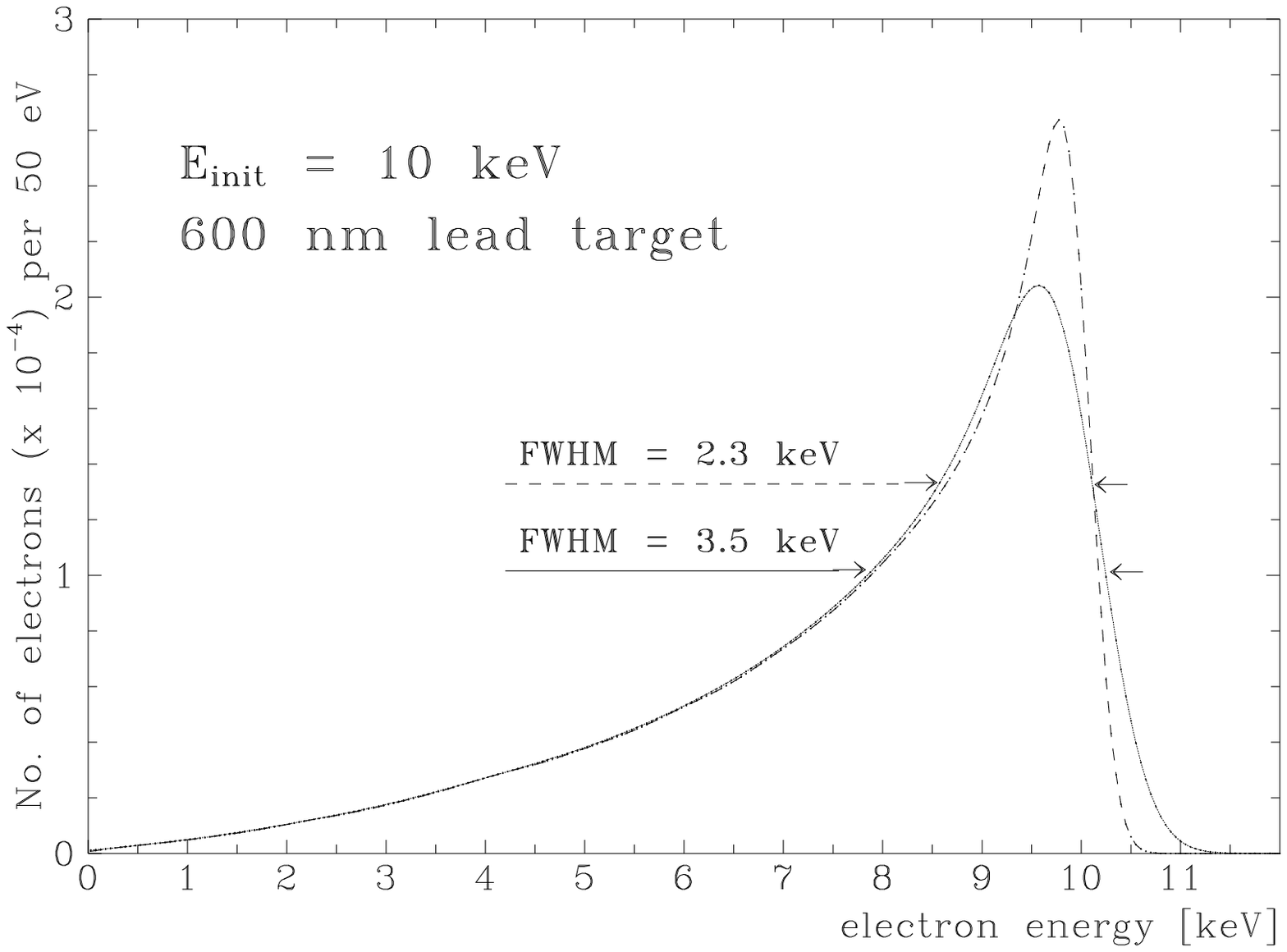}} \vspace{4ex}
\begin{figure}[h]
\protect\caption[fig3]{The convolution of the electron spectrometer response
   function with the Monte Carlo calculated energy spectrum of electrons 
   emitted from the 600~nm lead target (see text).
   The initial energy of electrons emitted from the sources in 
   the volume of lead target is 10 keV. The 
   dashed and full curves correspond to the 
   convolution of given energy distribution with the Gaussian  
   curves of FWHM = 0.5 and 1.0 keV, respectively.}
\label{f:tonda}
\end{figure}
\section{Conclusion}
     We have calculated the first internal conversion coefficients for
SH elements. These theoretical data should facilitate
planning of the heavy ion experiments aimed at production of the
SH nuclei as well as interpretation of the electron spectra
taken with future spectrometers. Via Monte
Carlo simulations we have demonstrated  that the electron 
scattering and energy losses
within today's reaction targets do not prevent the conversion
electron spectroscopy down to 10 keV.

     The extensive tables of the subshell and total ICC for 
104$\leq$Z$\leq$126 were deposited at LANL preprint server \cite{Ry00}.
For cases of special interest,
we can calculate the necessary ICC directly upon request.
\vspace{3ex} \\
{\bf Acknowledgement} ---\ This work was 
 supported by the Grant Agency of the
Czech Republic under contract No. 202/00/1625.
\vspace{3ex} \\
%
%

\end{document}